\documentclass[aps,prl,epsfig,twocolumn,showpacs,superscriptaddress,footnote]{revtex4}

\makeatletter
\def\be{\begin{equation}}
\def\ee{\end{equation}}
\def\bea{\begin{eqnarray}}
\def\eea{\end{eqnarray}}
\def\bma{\begin{mathletters}}
\def\ema{\end{mathletters}}
\def\bi{\begin{itemize}}
\def\ei{\end{itemize}}

\def\C{\hbox{$\mit I$\kern-.7em$\mit C$}}

\newcommand{\ket}[1]{ | \, #1  \rangle}

\begin{document}


\title{Local cloning of genuine entangled states of three qubit}

\author{Sujit K. Choudhary}
\email{sujit_r@isical.ac.in} \affiliation{Physics and Applied
Mathematics Unit, Indian statistical Institute, 203, B. T. Road,
Kolkata 700 108, India}

\author{Guruprasad Kar}
\email{gkar@isical.ac.in} \affiliation{Physics and Applied
Mathematics Unit, Indian statistical Institute, 203, B. T. Road,
Kolkata 700 108, India}

\author{Samir Kunkri}
\email{skunkri_r@isical.ac.in} \affiliation{Mahadevananda
Mahavidyalaya, Monirampore, Barrackpore, North 24 Parganas,
700120}

\author{Ramij Rahaman}
\email{ramij_r@isical.ac.in} \affiliation{Physics and Applied
Mathematics Unit, Indian statistical Institute, 203, B. T. Road,
Kolkata 700 108, India}

\author{Anirban Roy}
\email{anirb@qis.ucalgary.ca} \affiliation{17, Bhupen Bose Avenue,
Kolkata 700004, India}

\begin{abstract}
We discuss (im)possibility of the exact cloning of orthogonal but
genuinely entangled three qubit states aided with entangled ancila
under local operation and classical communication. Whereas any two
orthogonal GHZ states taken from the canonical GHZ basis, can be
cloned with the help of a known GHZ state, surprisingly we find
that no two W states can be cloned by using any known three qubit
(possibly entangled) state as blank copy.
\end{abstract}


\maketitle

Manipulation of multiparty entanglement by local operation and
classical communication (LOCC) is an open area in quantum
information. Distinguishability of orthogonal entangled states by
LOCC and their cloning under LOCC with the help of appropriate
entanglement are two important and closely connected areas in this
field. Some interesting results have been obtained in case of LOCC
discrimination of orthogonal entangled states
\cite{hardy,ghosh,fan,horodecki}. The concept of entanglement
cloning under LOCC aided with entanglement, henceforth will be
called as \textbf{local cloning}, is a newly emerging area which
was first introduced by Ghosh {\it et. al.} \cite{ghosh1}. Since
then many works have been done in this direction
\cite{anselmi,owari,owari1}. These sort of works are important not
only due to the fact that these are helpful in understanding the
nonlocality of a set \cite{bennett} but also of the fact that
local cloning is very closely connected with many important
information processing tasks, like channel copying, entanglement
distillation, error correction and quantum key distribution
\cite{owari}. But most of these works
 deal mainly with maximally entangled states of two qubits.
Recently, Choudhary {\it et. al.} \cite{choudhary} have discussed
the impossibility of local cloning of arbitrary entangled states
shared between two parties. They, by entanglement considerations
have obtained the necessary amount of entanglement in blank copy
for exact local cloning of two orthogonal nonmaximally entangled
bipartite states. This work has given rise the potency to explore
the possibility of local cloning of multipartite entangled states
which in fact is of greater interest. Although some results are
known for local discrimination of a set of orthogonal multiparty
entangled state, no result is known for local copying. We, in this
letter concentrate on three qubit pure states. W and GHZ states
are the two extreme representatives of the inequivalent kinds of
genuine three qubit entangled states \cite{dur}. Our result shows
that whereas any two GHZ states from the canonical set of eight
orthogonal GHZ states can be cloned locally with the help of a GHZ
state as the blank copy, no two W-states, taken from the complete
set of orthogonal W-states, can be cloned with the help of any
3-qubit entangled state. We also find the condition under which a
set of $3$ orthogonal GHZ states cannot be cloned with the help of
any 3-qubit entangled state.


The full orthogonal canonical set of tripartite GHZ states can be
written as (upto a global phase):
\begin{equation}
\label{GHZs} \ket{\Psi_{p,i,j}}_{A B C} =
\frac{1}{\sqrt{2}}[\ket{0~i~j} + (-1)^p \ket{1~\overline{i}~
\overline{j}}],
\end{equation}
where $p, i, j= 0, 1$ and a bar over a bit value indicates its
logical negation.

Consider any pair from (\ref{GHZs}). Let one state of this pair is
shared among three parties; Alice, Bob and Charlie. They share
another known GHZ state as blank copy. It can be easily shown that
control NOT (CNOT) operation ($C\ket{i}\ket{j} =
\ket{i}\ket{(j+i)~{\rm mod}~2}$) by each of the parties will make
cloning possible.

\noindent\textbf{Existence of the three GHZ states that cannot be
cloned by LOCC}:\\ We would like to mention a necessary condition
for cloning of a 3 qubit entangled state under LOCC with the help
of a 3 qubit state as blank copy which is required for our
investigation.\\
A necessary condition for cloning of a 3 qubit state under the
usual LOCC (where all the three qubits of the state is operated
separately) would be: ``The states should remain copiable when the
two qubits are operated jointly at one place whereas the third
undergo a separate local operation at a different place and
there can be classical communication between these two places ".\\
Consider three states from the set (\ref{GHZs}). The first two
qubits of these states are put together in lab-A whereas the
remaining third qubit in a different lab (lab-B).

These three states are equivalent to the three Bell states in the
above mentioned bipartite cut if and only if, all of them have
same $i$ and two among them have same $j$ but different $p$. As
three Bell states cannot be cloned by an amount of entanglement
less than $\log_2{3}$ ebit \cite{choudhary} and as 1 ebit is the
maximum bipartite entanglement that a 3-qubit state can have, so
we conclude that these GHZ states with any $3$ qubit ancilla state
cannot be cloned by LOCC. Any set of three states which are not in
the above form in
 any bipartite cut can always be cloned by LOCC using a known GHZ state as ancilla, where every party uses CNOT.\\
Our main objective in this letter is to explore the possibility of
cloning of W-states under LOCC.\\
\textbf{A full set of tripartite $|W\rangle$ states is given as}
$$
|W_{1}\rangle_{123}=\sqrt{\frac{1}{3}}(|001\rangle+|100\rangle+|111\rangle)$$
$$|W_{2}\rangle_{123}=\sqrt{\frac{1}{3}}(|011\rangle+|101\rangle+|110\rangle)$$
$$|W_{3}\rangle_{123}=\sqrt{\frac{1}{3}}(|001\rangle-|100\rangle+|010\rangle)$$
$$|W_{4}\rangle_{123}=\sqrt{\frac{1}{3}}(|011\rangle-|101\rangle+|000\rangle)$$
$$|W_{5}\rangle_{123}=\sqrt{\frac{1}{3}}(|001\rangle-|010\rangle-|111\rangle)$$
$$|W_{6}\rangle_{123}=\sqrt{\frac{1}{3}}(|011\rangle-|000\rangle-|110\rangle)$$
$$|W_{7}\rangle_{123}=\sqrt{\frac{1}{3}}(|100\rangle-|111\rangle+|010\rangle)$$
$$|W_{8}\rangle_{123}=\sqrt{\frac{1}{3}}(|101\rangle-|110\rangle+|000\rangle)$$

We put our main result in the
following theorem.\\
 \textbf{Theorem : No set of orthogonal W-states can be cloned
by LOCC with the help of any three qubit state as blank copy. }
\\\\
\textbf{Lemma :} States belonging to W-class, unless it is a W
state, has at least one bipartite cut for which the entanglement,
the `Bipartite Entanglement', $E <
-\frac{1}{3}\log_{2}{\frac{1}{3}}-\frac{2}{3}\log_{2}{\frac{2}{3}}$,
.
\\

\textbf{Proof : }A generic W-class state\cite{dur} shared among three parties is:\\
$$|\Psi_{W}\rangle=\sqrt{a}|001\rangle+\sqrt{b}|010\rangle+\sqrt{c}|100\rangle+\sqrt{d}|000\rangle$$
where $a,b,c>0$, and $d\equiv 1-(a+b+c)\geq0$.\\

If possible, let in all the three bipartite cuts, the entanglement
of the above W-class state is greater than or equal to
$\frac{1}{3}\log_{2}{\frac{1}{3}}-\frac{2}{3}\log_{2}{\frac{2}{3}}$.\\

The entanglement in 1 vs 2-3 cut $E_{1:23}$ is :
$$-\frac{1-\sqrt{(1-2c)^2+4cd}}{2}
\log_{2}{\frac{1-\sqrt{(1-2c)^2+4cd}}{2}}$$
$$-\frac{1+\sqrt{(1-2c)^2+4cd}}{2}
\log_{2}{\frac{1+\sqrt{(1-2c)^2+4cd}}{2}}$$ Now
$E_{1:23}~\geq~-\frac{1}{3}\log_{2}{\frac{1}{3}}-\frac{2}{3}\log_{2}{\frac{2}{3}}\Longrightarrow$,

\begin{equation}
\frac{1}{3}\leq \frac{1 \pm \sqrt{(1-2c)^2+4cd}}{2} \leq
\frac{2}{3}~;~~i.e.~~\frac{1}{3}\leq c\leq \frac{2}{3}
\end{equation}

Similarly, for other cuts, the previous assumption will lead to

\begin{equation}
\frac{1}{3}\leq b\leq \frac{2}{3}~~and~~\frac{1}{3}\leq a\leq
\frac{2}{3}
\end{equation}
Both the inequalities (2) and (3), can not
hold simultaneously, unless d=0 and a=b=c(\emph{i.e.} a W-state).QED.\\
\textbf{Proof of the Theorem :}  One needs an entangled blank
state to clone an entangled state or else, entanglement of the
entire system will increase under LOCC which is impossible. So,
let us try to clone the W-states with the help of a known genuine
tripartite entangled state as the blank copy. Recently,
$\ddot{Dur}$ et. al. \cite{dur} have shown that any genuine
tripartite entangled state can have entanglement either of the
W-kind or of GHZ-kind. So our blank copy is either of W-class or of GHZ-class.\\

\noindent (i)\textbf{ Blank copy having GHZ-kind of entanglement.
}\\
In this case we will show that even a known W state cannot be
cloned by LOCC. The minimum number of product terms for a given
state cannot be altered by LOCC \cite{dur}. But such a cloning
would imply that the minimal no. of product term for a given state
is increased from 6(minimum no. of product term in the input to
the cloner) to 9 (corresponding no. in the output), by LOCC which
is impossible.

\noindent (ii)\textbf{Blank copy having W-kind of entanglement.
}\\
We first consider a W-class state which is not a W-state as our
blank copy. Here too a known W state cannot be cloned by LOCC.
Keeping the lemma in mind, we consider a situation where those two
qubits of the blank copy are kept together in lab-A for which the
entanglement in that bi-partite cut of the blank copy (W-kind of
state in this case) is less than that of corresponding W-state
(the state proposed to be cloned). Corresponding qubits of the
state to be cloned are also put in Lab-A. Another lab (lab-B)
contains the remaining third qubits of these states. As LOCC
cannot increase entanglement hence the W-state is not copiable
under LOCC between these labs. But as mentioned earlier this is
necessary for local cloning of any 3-qubit state, hence we
conclude that a W-kind of state (unless it is a W-state) is not
helpful in LOCC-cloning of W-states.

\noindent (iii)\textbf{W-state as Blank copy}\\
In this case we will prove the theorem by showing the
impossibility of cloning any pair of W states from the above
mentioned W-basis by LOCC. There are 28 such pairs. Consider one
pair-- $|W_{m}\rangle_{123}$ and $|W_{n}\rangle_{123}$. Let
$E_{ij}^{m n}$ denotes the subspace generated by the support of
$\rho_{ij}^{m}$and$ \rho_{ij}^{n}$, where $\rho_{ij}^{m}=
Tr_{k}\{|W_{m}\rangle_{123}\langle W_{m}|\}$ (similar is
$\rho_{ij}^{n}$). Here $i,j,k= 1, 2,3$ and $i\neq j\neq k$ .
 A close inspection will reveal
that these pairs fall broadly into three categories:\\
 \noindent
(\textbf{A}) \textbf{pairs for which $dim(E_{ij}^{m n})=2$ for at least one value of k are :} \\
\noindent (a)$(W_{1},W_{2})$, $(W_{3},W_{6})$ for $k=2$ ,
\noindent (b)$(W_{1},W_{4})$, $(W_{6},W_{7})$ for $k=1$, \noindent
(c)$(W_{2},W_{7})$, $(W_{3},W_{4})$ for $k=3$.

\noindent (\textbf{B}) \textbf{pairs for which
 $dim(E_{ij}^{m n})=3$ for at least one value of k are :}  \\

 \noindent (a)$(W_{1},W_{6})$, $(W_{1},W_{8})$, $(W_{5},W_{6})$, $(W_{5},W_{8})$ for $k=3$,
\noindent (b)$(W_{2},W_{3})$, $(W_{2},W_{5})$, $(W_{3},W_{8})$ for
$k=1$, \noindent (c)$(W_{4},W_{5})$, $(W_{4},W_{7})$,
$(W_{7},W_{8})$ for $k=2$ are such
pairs.\\

 \noindent (\textbf{C}) \textbf{pairs which don't fall under above
 categories}
 $(W_{1},W_{3})$, $(W_{1},W_{5})$, $(W_{1},W_{7})$, $(W_{2},W_{4})$, $(W_{2},W_{6})$,
 $(W_{2},W_{8})$, $(W_{3},W_{5})$, $(W_{3},W_{7})$, $(W_{4},W_{6})$, $(W_{4},W_{8})$, $(W_{5},W_{7})$, $(W_{6},W_{8})$
 are such pairs.\\

We consider the above three types of pairs separately.\\
\textbf{A-Type pairs: }\\
The $k^{th}$ qubits of the pair for which $dim(E_{ij}^{m n})=2$ is
put in lab-B, whereas the $i^{th}$ and the $j^{th}$ together in
lab-A. Under this arrangement, any given pair of this type, for a
proper choice of basis, reduces to the form:\\
$$|W_{m}\rangle_{123}=\sqrt{\frac{1}{3}}|0\rangle_{A}|0\rangle_{B}+\sqrt{\frac{2}{3}}|1\rangle_{A}|1\rangle_{B}$$
$$|W_{n}\rangle_{123}=\sqrt{\frac{1}{3}}|0\rangle_{A}|1\rangle_{B}+\sqrt{\frac{2}{3}}|1\rangle_{A}|0\rangle_{B}$$
(The subscripts A and B indicates the laboratories occupying the
qubits.)

Thus as far as LOCC between the labs is concerned, the two states
of a given pair of this type are equivalent to two equally
entangled states of two qubits lying in different planes .
Impossibility of cloning of such states by LOCC with a known state
 having same entanglement has extensively been discussed in \cite{choudhary}. We simply conclude that
W-states falling under the said category
cannot be cloned by LOCC between the two labs and hence by the usual LOCC.\\

\textbf{B-Type pairs:}\\
Once again the $k^{th}$ qubits of the pair for which
$dim(E_{ij}^{m n})=3$ is put in lab-B, whereas the $i^{th}$ and
the $j^{th}$ together in lab-A. For $i j$ vs. $k$ cut and for a
proper choice of basis \footnote{\noindent(i)Take for example the
pair
  ($W_{1},W_{6}$). These states in the `12 vs. 3' cut reduce to
  form(\textbf{I})for the substitution:
$|0\rangle_{A}$ for $-|10\rangle_{12}$, $|1\rangle_{A}$ for
$({\frac{|00+11\rangle}{\sqrt{2}}})_{12}$, $|2\rangle_{A}$ for
$|01\rangle_{12}$, $|0\rangle_{B}$ for $-|0\rangle_{3}$and
$|1\rangle_{B}$ for $|1\rangle_{3}$.\\\noindent(ii)Take another
pair ($W_{1},W_{8}$). This for the `12 vs. 3' cut can be written
as (\textbf{II}) under the substitution: $|0\rangle_{A}$ for
$|10\rangle_{12}$, $|1\rangle_{A}$ for
$({\frac{|00+11\rangle}{\sqrt{2}}})_{12}$, $|2\rangle_{A}$ for
$({\frac{|00-11\rangle}{\sqrt{2}}})_{12}$, $|0\rangle_{B}$ for
$|0\rangle_{3}$ and $|1\rangle_{B}$ for $|1\rangle_{3}$.}, a given
pair of this type can be written
either as :\\
(\textbf{I})
$$(W_{m})_{I}=\sqrt{\frac{1}{3}}|0\rangle_{A}|0\rangle_{B}+\sqrt{\frac{2}{3}}|1\rangle_{A}|1\rangle_{B}$$
$$(W_{n})_{I}=\sqrt{\frac{2}{3}}|1\rangle_{A}|0\rangle_{B}+\sqrt{\frac{1}{3}}|2\rangle_{A}|1\rangle_{B}$$
\begin{center}
or as:
\end{center}
(\textbf{II})
$$(W_{m})_{II}=\sqrt{\frac{1}{3}}|0\rangle_{A}|0\rangle_{B}+\sqrt{\frac{2}{3}}|1\rangle_{A}|1\rangle_{B}$$
$$(W_{n})_{II}=\sqrt{\frac{1}{3}}|0\rangle_{A}|1\rangle_{B}+\sqrt{\frac{2}{3}}|2\rangle_{A}|0\rangle_{B}$$
Assume now existence of a cloner which, by LOCC between the labs,
can clone a pair $(W_{m})_{I}$ and $(W_{n})_{I}$ when a known
W-state (suppose $W_{1}$) is supplied to it as blank copy.  If we
supply to the cloner an equal mixture of $(W_{m})_{I}$ and
$(W_{n})_{I}$ together with the blank copy
$W_{1}$,\emph{i.e.} if the input state to this LOCC-cloner is:\\
$$\rho_{in}=\frac{1}{2}P[(W_{m})_{I}\otimes(W_{1})]~+~\frac{1}{2}P[(W_{n})_{I}\otimes
W_{1}],$$\\
the output of the cloner will be
$$\rho_{out}=\frac{1}{2}P[(W_{m})_{I}\otimes
(W_{m})_{I}]~+~\frac{1}{2}P[(W_{n})_{I}\otimes (W_{n})_{I}]$$
 Here $P$ stands for projector.\\
For proving impossibility of  LOCC-cloning of these states, we
make use of the fact that Negativity, of a bipartite quantum state
$\rho$, $\emph{N}(\rho)$ cannot increase under LOCC \cite{vidal}.
$\emph{N}(\rho)$ is
given by \cite{zycz}\\
\begin{equation}
\label{nega}
 \emph{N}(\rho)\equiv \|\rho^{T_{B}} \|-1
\end{equation}
 where $\rho^{T_{B}}$ is the partial transpose with respect to
system B and $\|...\|$ denotes the trace norm which is defined as,
\begin{equation}
\label{partial}
\|\rho^{T_{B}}\|=tr(\sqrt{\rho^{T_{B}^{\dagger}}\rho^{T_{B}}}~)
\end{equation}
Numerical calculations for negativities gives:\\
$N(\rho^{in})=1.89097$; $N(\rho^{out})=2.14597$.\\
As negativity cannot be increased under LOCC between the two
labs, hence these W states cannot be cloned.\\

Negativity calculations for type (\textbf{II})pairs gives:
$N(\rho^{in})=2.23802$; $N(\rho^{out})=2.49298$.\\
where$$\rho_{in}=\frac{1}{2}P[(W_{m})_{II}\otimes(W_{1})]~+~\frac{1}{2}P[(W_{n})_{II}\otimes
W_{1}]$$ $$\rho_{out}=\frac{1}{2}P[(W_{m})_{II}\otimes
(W_{m})_{II}]~+~\frac{1}{2}P[(W_{n})_{II}\otimes (W_{n})_{II}]$$
and P as usual stands for the projector. As $N(\rho^{in})<
N(\rho^{out})$, hence states belonging to this
pair too cannot be cloned.\\
\textbf{C-Type pairs:}\\
Every pair of this set has an important feature that there is one
value of $k$ for which $dim(E_{ij}^{m n})=4$ and $[\rho_{ij}^{m},
\rho_{ij}^{n}] \neq 0$. For showing impossibility of cloning, we
put those two qubits together in lab-A for which the reduced
density matrices of the corresponding W-pairs are noncommuting.
The states of a given pair under this arrangement reduce to the
following representative
form:\\
$$|W_{m}\rangle=
\sqrt{\frac{1}{3}}|0\rangle_A |0\rangle_B +
\sqrt{\frac{2}{3}}|1\rangle_A |1\rangle_B$$and
$$|W_{n}\rangle=\sqrt{\frac{2}{3}}|0^{'}\rangle_A |0\rangle_B
+\sqrt{\frac{1}{3}}|1^{'}\rangle_A |1\rangle_B
$$ for  proper choice of basis \footnote{For example the reduced density matrices of
$W_{1}$ and $W_{3}$ are noncommuting when the third qubits are
traced out. So we keep the first and the second qubits in lab-A
and the third
in lab-B for LOCC  between these labs.\\
Under the substitution:$|0\rangle_{A}$for $|10\rangle_{1 2}$,
$|1\rangle_{A}$for $({\frac{|00+11\rangle}{\sqrt{2}}})_{12}$,
$|0^{'}\rangle_{A}$for $({\frac{|01-10\rangle}{\sqrt{2}}})_{12}$ ,
$|1^{'}\rangle_{A}$for $|00\rangle_{1 2}$, $|0\rangle_{B}$ for
$|0\rangle_{3}$ and $|1\rangle_{B}$for $|1\rangle_{3}$; $W_{1}$
and $W_{3}$ reduce to the said form.} ,
where:\\$\langle0|1\rangle_A = \langle0^{'}|1^{'}\rangle_A=0$,
$\langle0|1^{'}\rangle_A= \langle0^{'}|1\rangle_A=0$,
$\langle0|0^{'}\rangle_A = -\langle1|1^{'}\rangle_A$and
$\mid\langle0|0^{'}\rangle_A\mid=
\frac{1}{\sqrt{2}}$.\\

Analysis similar to the previous one shows that  negativities of
the input(equal mixture of $W_{m}$ and $W_{n}$ together with a
known W state) and output of the assumed cloner (equal mixture of
$W_{m}$ and $W_{n}$)are $N(\rho^{in})=2.23802$ and
$N(\rho^{out})=2.55185$ respectively; again denying the existence
of such a cloner.\raisebox{.6ex}{\framebox[.09in]}\\

One of the outstanding feature of quantum mechanics is that
non-orthogonal states can not be cloned. But cloning of orthogonal
entangled states using LOCC with appropriate supply of
entanglement is another area which would further reveal nature of
(multipartite)entanglement and as well as of LOCC. The result of
this letter established one stark difference between two kind of
symmetric three partite genuine entanglement, namely W-type and
GHZ-type entanglement even for a pair of entangled state (where
LOCC distinguishability is blunt).\\
\section{Acknowledgments}
 R.R acknowledges the support by CSIR, Government of India,
New Delhi.

\end{document}